\documentclass[10pt]{article}
\usepackage{amssymb}
\usepackage{graphicx}
\usepackage{fancyhdr}

\newcommand{\beq}{\begin{equation}}
\newcommand{\eeq}{\end{equation}}
\newcommand{\bear}{\begin{eqnarray}}
\newcommand{\eear}{\end{eqnarray}}
\newcommand{\bearn}{\begin{eqnarray*}}
\newcommand{\eearn}{\end{eqnarray*}}

\begin{document}

\input epsf

\begin{center}
{\bf{\Large{ Higher-dimensional Temperley-Lieb algebras }}}
\end{center}

\begin{center}
{\bf{\large{\bf Marcos Alvarez\qquad Paul P. Martin}}}
\end{center}

\begin{center} 
Centre for Mathematical Science, City University \\
Northampton Square, London EC1V 0HB, UK \\
{\tt{e-mail: m.alvarez@city.ac.uk\qquad p.p.martin@city.ac.uk}}
\end{center}

\begin{abstract}

A category which generalises to higher dimensions many of the features of 
the Temperley-Lieb category is introduced. 

\end{abstract}

{\hskip 10pt{MSC-class: 81R50 (primary), 82B20 (secondary).}}


\def\bb#1{{\mathbb{#1}}}

\def\comment#1
{\bigskip\hrule\medskip{\begin{quotation}\noindent\bf{#1}\end{quotation}\medskip\hrule\bigskip}}


\newcount\defnumber
\newcount\propnumber
\newcount\theonumber
\newcount\notnumber
\newcount\lemmanumber
\newcount\fignumber
\newcount\anything

\def\definition#1#2{\medskip\global\advance\anything by 1 
\noindent{\bf Definition \the\anything\hskip 5pt} 
{\it #2}\smallskip} 

\def\proposition#1#2{\medskip\global\advance\anything by 1 
\noindent{\bf Proposition \the\anything\hskip 5pt} 
{\it #2}\smallskip} 

\def\theorem#1#2{\medskip\global\advance\anything by 1 
\noindent{\bf Theorem \the\anything\hskip 5pt} 
{\it #2}\smallskip} 


\def\lemma#1#2{\medskip\global\advance\anything by 1 
\noindent{\bf Lemma \the\anything\hskip 5pt} 
{\it #2}\smallskip} 


\def\num{(\thesection.\the\eqnum)}
\def\appnum{({\sl{A}}\the\appnumber.\the\aeqnum)}
\def\deflabel{\thesection.\the\defnumber}
\def\proplabel{\thesection.\the\propnumber}
\def\theolabel{\thesection.\the\theonumber}
\def\notlabel{\thesection.\the\notnumber}
\def\lemlabel{\thesection.\the\lemmanumber}

\def\anylabel{\the\anything}

\def\nlabel{\edef}


\def\ie{{\it{i.e.\/}}}
\def\eg{{\it{e.g.\/}}}
\def\N{{\bb{N}}}
\def\Z{{\bb{Z}}}
\def\E{{\bb{E}}}
\def\R{{\bb{R}}}
\def\G{{\cal{G}}}
\def\nobub{{\sf{R}}}
\def\mapd{{\bb{D}}}
\def\suchthat{{\,\,|\,\,}}
\def\at#1{{\big\vert_{#1}}}

\def\S{{S}}
\def\C{{\cal{C}}}
\def\K{{\cal{K}}}

\def\block{{$\Box$}}

\def\isotopy{{{\mathfrak{i}}}}
\def\striso{{{\mathfrak{si}}}}
\def\relrone{{{\mathfrak{r}_1}}}
\def\relr{{{\mathfrak{r}}}}
\def\het{{{\mathfrak{h}}}}
\def\strhet{{{\mathfrak{sh}}}}

\def\here{{\vskip 1cm\hrule\vskip .5cm REVISE HERE \vskip
.5cm\hrule\vskip 1cm}}
\def\incomplete{{\vskip 1cm\hrule\vskip .5cm INCOMPLETE \vskip
.5cm\hrule\vskip 1cm}}

\def\np{Nucl.~Phys.}
\def\pl{Phys.~Lett.}
\def\pr{Phys.~Rev.}
\def\cmp{Commun.~Math.~Phys.}
\def\prs{Proc.~Roy.~Soc.}



\section{Introduction}
\setcounter{equation}{0}

The Temperley-Lieb category provides a useful tool in computational 2d 
lattice statistical mechanics \cite{baxterbook}.
Its representation theory is the interlocutor between several
different `equivalent' lattice models (Potts, IRF, 6-vertex, etc.) \cite{baxterbook, ppm}, 
is universal among an important class of solutions to integrability
conditions, and provides the invariant theory for $U_qsl_2$ \cite{jimbo, kassel}.
Its categorical structure means that its representation theory can be
analysed in great generality \cite{ppm, adamovich} (it is universal among monoidal dual
categories with certain natural properties, and was the starting point
for Khovanov homology \cite{khovanov}). It is also important in a number of other
areas of Mathematics and Physics \cite{kaufman, koo}.

The generalisation to `higher dimensions' (in the mean-field sense of
every lattice point being a neighbour) is the partition category \cite{ppmtl},
but it is natural also to seek a 3d version. A 3d lattice subalgebra
of the partition algebra was studied by Dasmahapatra and Martin \cite{dasmahapatra}
but it has very few of the beautiful properties of the Temperley-Lieb and
partition categories (cf. Baxter-Bazhanov tetrahedra for example \cite{bb, bbtetra}),
and essentially no progress has been made in this direction. Here we
describe a generalisation of the full Temperley-Lieb category to 3d that is 
closer in spirit to the cobordism category of Topological Quantum Field Theory \cite{baez, laures}
and to Lattice Gauge Theories (\cite{ppm}, page 278).
The problem with such a construction is that there are significant 
initial problems with well-definedness, but the payoff is potential access to generalisations of several of
the structures mentioned above. In this note we solve the well-definedness problem, and the core 
enumeration problem in using the resultant algebras. Applications will be discussed elsewhere.

\section{Concrete diagram categories}
\setcounter{equation}{0}
{\global\propnumber=0}
{\global\defnumber=0}
{\global\lemmanumber=0}
{\global\theonumber=0}
{\global\notnumber=0}

We begin by selecting a direction in $d$-dimensional Euclidean space $\E^d$ which 
we call ``time'', coordinatised by a real number $t$. For fixed $d\in\N$ and $t\geq 0$ we define 
$E_t=\R^{d-1}\times [0,t]$. If $D$ and $E$ are subsets of $\E^d$, we write $C_E(D)$ for the number of 
connected components of $E\setminus D$ in the Euclidean topology.

\definition{condiagppm}{
The set $S^{d,t}$ of {\rm concrete diagrams of duration $t$} is the set of compact subsets of $E_t$ such that
$D\in S^{d,t}$ if and only if, for all $D'$ obtained from $D$ by removing a single point, we have  
$C_{E_t}(D')=C_{E_t}(D)-1$. Then $E_t$ is called a {\rm universe} for $D$
Write $S^d$ for $\cup_{t\geq 0} S^{d,t}$ and $S^{d-1}_{[]}$ 
for the set of concrete diagrams in $S^d$ with $t=0$.}
\nlabel\concdiagppm{\anylabel} 

\begin{figure}
  \begin{center}
    \begin{minipage}[c]{.58\linewidth}
      \epsffile{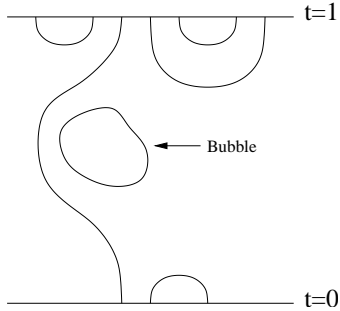}
    \end{minipage}\hfill
    \begin{minipage}[c]{0.4\linewidth} 
		{\caption{A concrete diagram in $S^{2,1}$ having a bubble: \newline $b(D)=1$, $g(D)=0$, $|D|=6$,		
						\newline $\chi(D)=|D|-b(D)=5$.}}
    \end{minipage}
  \end{center}
\end{figure}

For $d=2$, concrete diagrams coincide with concrete Temperley-Lieb diagrams, and in this sense our 
construction provides a ``higher-dimensional'' generalisation. 
For $d=3$, a concrete diagram can be thought of as consisting of a number of 
embedded submanifolds whose boundaries all lie on either of two parallel planes of constant time. 
See Figures 1 and 2 for examples.

\begin{figure}
\centering
\epsffile{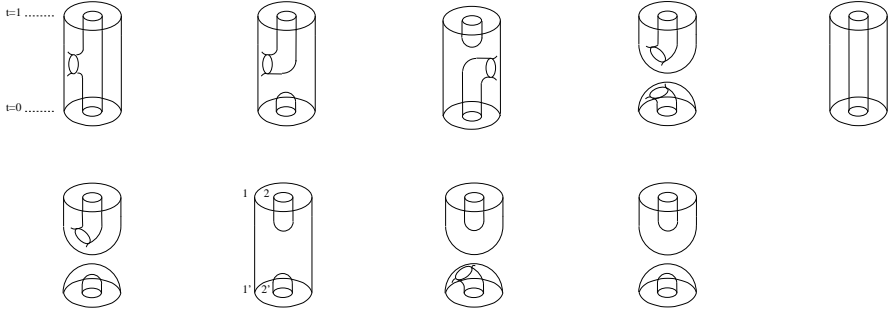}
\caption{Representations of several concrete diagrams in $S^{3,1}$ whose boundaries are two pairs of concentric circles. One concrete diagram has its boundary loops labelled 1 ,2, 1' and 2' for a later purpose.}
\end{figure}

A {\it component} of $D\in S^{d,t}$ is any point $x\in D$ together with the maximal subset
$X\ni x$ such that $C_{E_t}(D\setminus X)=C_{E_t}(D\setminus \{x\})$.
A {\it bubble} is a component which does not intersect
the boundaries of $E_t$. A component $c$ has a {\it handle} if it properly contains a non-contractible loop 
not homologous in $c$ to a subset of the boundary of $E_t$. Write $|D|$ for the number of components of a
concrete diagram $D$, and $b(D)$ for the number of bubbles.

The {\it number of handles} of a concrete diagram $D\in S^3$ is the genus of $D$, written 
$g(D)$. We will write $\chi(D)$ for the Euler number of $D$ \cite{armstrong, mendels}.

\definition{sffppm}{
Let $F,F'\subset \R^{d-1}$. Then $S^{d,t}[F,F']$ is the subset of $S^{d,t}$ of concrete diagrams
which intersects the t-boundary of $E_t$ in $F$ and the 0-boundary in $F'$. 
Set $S^d[F,F']=\cup_{t\geq 0} S^{d,t}[F,F']$. The {\rm boundary configuration} of 
$D\in S^d[F,F']$ is the ordered pair $(F,F')$. }
\nlabel\sffppm{\anylabel} 

(N.B.: $S^d[F,F']=\emptyset$ unless $F,F'\in S^{d-1}_{[]}$. In 
practice we shall restrict attention to cases in which $F$ and $F'$ are unions of $(d-2)$-spheres.)

For $T$ a set, let $\wp(T)$ be the set of partitions of $T$ and ${\cal P}(T)$ the power set of $T$.  
For $U$ and $V$ sets and $u\in\wp(U)$, $v\in\wp(V)$ we write $u\ast v$ for the partition
algebra composition of $u$ with $v$ (\cite{ppmsaleur} p.158, or more generally the $Ag$ product in 
\cite{ppmtl}, p.77).

For $D\in S^d[F,F'] $ define its {\it connectivity} $p(D)$ as the partition of the components 
(with respect to the obvious extension of the $\E^{d-1}$ topology) of $F\dot\cup F'$, such that $a,b$ are
in the same part if they are in the same component of $D$. 
For example, for $D$ the concrete diagram with labelled boundaries in Figure 2, the connectivity 
is $p(D)=\{\{1,1'\},\{2\},\{2'\}\}$.

For fixed $d\in\N$ define the map $\nobub:S^d\to S^d$
by $\nobub(D)$ being $D$ with all bubbles removed.  
Define $S^d_{min}\subset S^d$ as the subset of elements with no bubbles and no handles.
The elements of $S^d_{min}$ are called {\it minimal}. 
For example, all the diagrams shown in Figure 2 are minimal, whereas the diagram of Figure 1 is not, as it 
contains a bubble. 

For any two sets $A$ and $B$, define the symmetric difference 
$A\wedge B:=(A\cup B)\setminus(A\cap B)$. For $A,B\in S^d$ we write 
$A\overline{\wedge}B$ for the closure of $A\wedge B$ in the Euclidean metric 
topology. Two concrete diagrams $D$ and $D'$ are {\it $\wedge$-composable} if they have a 
universe in common and $D\cap D'$ is a finite union of disjoint 
closed $(d-1)$-balls. 
It can then be shown that \cite{AM}

\lemma{wed}{
If $D$ and $D'$ are $\wedge$-composable then $D\overline{\wedge}D'\in S^d$. }\block
\nlabel\wed{\anylabel} 

For $B\in S^d$, define $dom B$ as the set of all $A\in S^d$ which are $\wedge$-composable
with $B$. Then we define the map $\delta_B: dom B \to S^d$ by $A \mapsto  A\overline{\wedge}B$.

For any point $P\in\E^d$ with time coordinate $t_P$, the {\it time translate} $P_\tau$ is the point in $\E^d$
with the same projection on the $t=0$ subspace as $P$ but with time coordinate $t_P+\tau$. For any 
$Z\subset\E^d$, we define $Z_\tau$ by $P\in Z_\tau\Leftrightarrow P_{-\tau}\in Z$. The 
following diagram composition Lemma follows now from Definition \concdiagppm:

\lemma{conc}{
If $A\in S^{d,t}[F',F]$ and $B\in S^{d,\tau}[F,F'']$, then $A\circ B:=A_{\tau}\cup B$ is in $S^d[F',F'']$. }\block
\nlabel\conc{\anylabel} 

Let us now define the triple $\C^d=(S_{[]}^{d-1}\>,\> \hom(-,-),\>\circ\>)$
consisting of the ``object set'' $S_{[]}^{d-1}$; 
and for each pair of objects $E,F\in S_{[]}^{d-1}$ the collection of ``morphisms'' 
$\hom(E,F)=S^d[E,F]$; and composition of
morphisms defined by $\circ$-composition of concrete diagrams. The morphism in $\hom(F,F)$ of duration 
one whose sections of constant time are time translates of $F$ will be denoted ${\mathfrak I}_F$.

\theorem{cdcat}{
$\C^d$ is a category.}

{\sl Proof:} We require to show associativity of $\circ$, and existence of identity element in each
$hom(F,F)$. The former is clear, the latter is the concrete diagram of duration zero. \block
\nlabel\cdcat{\anylabel} 

Let $A\in S^3[F',F]$ and $B\in S^3[F,F'']$. Then
\beq
\chi(A\circ B)=\chi(A)+\chi(B),
\label{eq:euleradd}
\eeq
\nlabel\chicirc{\lemlabel}\noindent
(which follows directly from the definition of $\chi(A)$ as an alternating sum of
Betti numbers, or as a sum of Morse indices \cite{Milnor}, of $A$) and
\beq
g(A\circ B) =g(A)+g(B)+|A\circ B| -|A|-|B|+|F|.
\label{eq:genusformula}
\eeq
We will later need other results that follow easily from Eq. (\ref{eq:euleradd}) which we give
without proof. Here we assume that all $\circ$-compositions are defined; 
$D$ is a minimal concrete diagram with the same connectivity as $A\circ B$, and 
$G$ is a minimal concrete diagram with the same connectivity as $B\circ C$:
\beq
\begin{array}{ccc}
&g(A\circ B)+g(D\circ C) = g(B\circ C)+g(A\circ G),& \\
&b(A\circ B)+b(D\circ C) = b(B\circ C)+b(A\circ G).&
\end{array}
\label{eq:bandg}
\eeq

\section{Equivalence relations on $S^3$}
\setcounter{equation}{0}
{\global\propnumber=0}
{\global\defnumber=0}
{\global\lemmanumber=0}
{\global\theonumber=0}
{\global\notnumber=0}

In this section we set $d=3$. We will write $\isotopy$ for the usual relation of isotopy 
(\ie, a continuous, one-parameter family of homeomorphisms {\cite{Moise}}). 
If $\mathfrak j$ is a specific isotopy with parameter $s\in[0,1]$ and $A$ 
is a concrete diagram, we write ${\mathfrak j}_u(A)$  for the image of $A$ under 
${\mathfrak j}$ at $s=u$. In particular, ${\mathfrak j}_0(A)=A$ for all $A$.
Two concrete diagrams $A$ and $B$ are {\it isotopic} iff there is an isotopy $\mathfrak j$ 
such that ${\mathfrak j}_s(A)$ is a concrete diagram for all $s$, and
${\mathfrak j}_1(A)=B$.

This defines the isotopy relation $\isotopy$ in the set of concrete diagrams. It is obviously an 
equivalence relation.

Let $A$ and $B$ be concrete diagrams. If $A$ and $B$ are isotopic by an isotopy ${\mathfrak j}$
such that the boundaries of ${\mathfrak j}_s(A)$ are time translates of the boundaries of $A$ for all $s$
we say that they are {\it strongly isotopic}. 

For example, the isotopy whose action on a point in $\E^3$ with coordinates $(x,y,t)$ is
${\mathfrak j}_s(x,y,t)=(x,y,(1+s)t)$ is strong. Strong isotopy is clearly an equivalence relation, which we 
denote $\striso$.

\definition{relations}{
\\ 1. For any $A,B\in S^3$, define relation $\relrone$ by $A\relrone B$ if there is a torus $T$ such that 
$A\cap T$ is a disk, and $B=\delta_T A$. Relation $\relr$ is the transitive closure of $\relrone$.
\\2. Relation $\strhet$, called {\rm strong heterotopy}, is the reflexive, symmetric and transitive closure of 
$\relrone$ and $\striso$.
\\3. Relation $\het$, called {\rm heterotopy}, is the reflexive, symmetric and transitive closure of 
$\relrone$ and $\isotopy$. }
\nlabel\relations{\anylabel} 

The cosets $S^3[F,F']/\strhet$ and $S^2[F,F']/\isotopy$ (Temperley-Lieb diagrams) are infinite owing to the 
possible presence of bubbles or handles. But $S^2_{min}[F,F']/\isotopy$ is finite. We will show in 
section~\ref{scon} that $S^3_{min}[F,F']/\strhet$ is finite.

The following Lemma follows immediately from the previous definitions:

\lemma{bubhet}{
If $A\het A'$ then $b(A)=b(A')$. If $A\strhet A'$ then $p(A)=p(A')$. }\block
\nlabel\bubhet{\anylabel} 

\definition{hetclass}{
For $D\in S_{min}^3$, write $[D]_\het$ for the restriction of the $\het$-class of $D$ to
$S_{min}^3$, \ie, $[D]_\het=\{C\in S_{min}^3 \suchthat C\het D\}$. 
Write $\S_\het^3$ for the set of $\het$-classes in $S_{min}^3$ and $\S_\het^3[F,F']$ for the
set of $\het$-classes in $S_{min}^3[F,F']$. Similarly, write $[D]_\strhet$ for 
the restriction of the $\strhet$-class of $D$ to $S_{min}^3$, \ie,
$[D]_\strhet=\{C\in S_{min}^3 \suchthat C\strhet D\}$, and 
$\S_\strhet^3$ for the set of $\strhet$-classes in $S_{min}^3$, and $\S_\strhet^3[F,F']$ for the
set of $\strhet$-classes in $S_{min}^3[F,F']$.}
\nlabel\hetclass{\anylabel} 

By the handle decomposition theorem \cite{Milnor}:

\proposition{minrep}{
For every $A\in S^3$ there exists $D\in S_{min}^3$ such that $D\relr \nobub(A)$. \block}
\nlabel\minrep{\anylabel} 

\definition{mapdr}{
Define $\mapd_r:S^3 \to  {\cal P}(S_{min}^3)$ by 
$A \mapsto \{D\in S_{min}^3 \suchthat D\relr\nobub(A)\}$.}
\nlabel\mapdr{\anylabel} 

The following Proposition follows from Proposition \minrep\ and Definition \relations.

\proposition{mapdvsh}{
For each $A\in S^3$  there is $D\in S_{min}^3$ such that $\mapd_r(A)\subseteq[D]_\strhet$.
Moreover, if $\mapd_r(A)\subseteq[D]_\strhet$ and $\mapd_r(A)\subseteq[D']_\strhet$ then
$[D]_\strhet=[D']_\strhet$. }\block
\nlabel\mapdvsh{\anylabel} 
%

Define $\mapd_\het(A)$ and $\mapd_\strhet(A)$ as, respectively, the $\het$- and $\strhet$-class 
in $S^3_{min}$ containing $\mapd_r(A)$. 

\definition{redmap}{
For $\K$ a ring and $p,q\in\K$ define the {\rm reduction maps} $\mu$ and $\nu$ as
\bearn\centering
\begin{array}{rclccrcl}
\mu: S^3 &\to & \K \S_\het^3    &  &    &      \nu: S^3 &\to & \K \S_\strhet^3       \\
A &\mapsto & p^{g(A)}q^{b(A)}\mapd_\het(A)  &&& A &\mapsto & p^{g(A)}q^{b(A)}\mapd_\strhet(A)
\end{array}
\eearn
and extend linearly to $\K \S^3$ in each case.}
\nlabel\redmap{\anylabel} 

If $A$ and $B$ are $\circ$-composable, then any $A'\strhet A$ and $B'\strhet B$ are also 
$\circ$-composable. The next Lemma follows from noticing that the two strong heterotopies relating 
$A$ to $A'$ and $B$ to $B'$ combine into a single one that relates $A\circ B$ to $A'\circ B'$:

\lemma{shsh}{
Let $A$ and $B$ are $\circ$-composable. Let $A'\strhet A$ and $B'\strhet B$. Then
$(A\circ B)\strhet(A'\circ B')$. \block}
\nlabel\shsh{\anylabel} 

Therefore we can extend $\circ$-composability to $\strhet$-classes 
in a well-defined way by $\circ$-composing representatives.

\definition{circsh}{
Let $A,B$ be $\circ$-composable. Then $[A]_\strhet \bullet [B]_\strhet := \nu(A\circ B)$.}
\nlabel\circsh{\anylabel} 

\theorem{catsh}{
The triple $\C_{sh}=(S^2_{[]}\>,\> \K\S^3_\strhet[-,-],\> \bullet\>)$
is a category whose morphisms are $\strhet$-classes of concrete diagrams.}

{\sl Proof: } i) Associativity of $\bullet$-composition of $\strhet$-classes follows from 
associativity of $\circ$-composition of concrete diagrams. ii) The unit in $\S^3_\strhet[F,F]$
is the $\strhet$-class of the diagram of zero duration in $S^3[F,F]$. \block

%

\proposition{hettostrhet}{
Let $A,B\in S^d[F,F']$ and $A\het B$. Then there exist $L\in S^3_{min}[F,F]$ and $R\in S^3_{min}[F',F']$ 
such that $L\het {\mathfrak I}_F$, $R\het {\mathfrak I}_{F'}$, and $A\strhet(L\circ B\circ R)$.}

{\sl Proof:} If $A\strhet B$ then take $L={\mathfrak I}_F$ and $R={\mathfrak I}_{F'}$. Otherwise 
the $\het$-relation between 
$A$ and $B$ contains isotopies in a neighbourhood of the boundary of $A$. Extend those isotopies to
a neighbourhood of ${\mathfrak I}_F$ and ${\mathfrak I}_{F'}$ in 
${\mathfrak I}_F\circ A\circ{\mathfrak I}_{F'}$
to obtain $L\circ B\circ R$. The $L$ and $R$ so defined are clearly $\het$-related to 
${\mathfrak I}_F$ and ${\mathfrak I}_{F'}$ respectively (in fact, isotopic). \block
\nlabel\hettostrhet{\anylabel} 

By a routine check that the group axioms are satisfied, we have

\proposition{idclasses}{
The $\strhet$-classes inside $[{ \mathfrak I}_F]_\het$ in $S_{min}^3[F,F]$ form a group 
$\Pi_F$ under $\bullet$-composition, with unit $[{ \mathfrak I}_F]_\strhet$. \block}
\nlabel\idclasses{\anylabel} 

\definition{sumx}{
For a finite $X\subset S^d_{min}$, define $\sigma_{\scriptscriptstyle X}=\sum_{A\in X} A\in\K S_{min}^d$.}
\nlabel\sumx{\anylabel} 

\section{$\het$-classes and connectivity.}\label{scon}
\setcounter{equation}{0}
{\global\propnumber=0}
{\global\defnumber=0}
{\global\lemmanumber=0}
{\global\theonumber=0}
{\global\notnumber=0}

In this Section we show that the $S^3[F,F']/\strhet$ is finite.

\definition{neighbours}{
Let $c$ and $d$ be components of $A\in S^d$. Then $c$ and $d$ are {\rm neighbours} if there is a
path connecting $c$ to $d$ not intersecting any other component of $A$. Let $c$ and $d$ be neighbours in 
$A$ with path $P$, and $e$ a surface obtained from $c$ and $d$ by removing a disk from each and joining the 
edges with a cylindrical thickening of $P$ not intersecting any other component of $A$. Then $e$ is a 
{\rm bridging} of $c$ and $d$ and the cylinder is a {\rm bridge} connecting $c$ and $d$.}
\nlabel\neighbours{\anylabel} 

\theorem{bigtheorem}{
Let $A,B\in S^3_{min}[F,F']$. Then $A\strhet B$ iff $p(A)=p(B)$. \block}
\nlabel\bigtheorem{\anylabel} 

Necessity follows from the definition of $\strhet$. A complete proof of sufficiency will be given in \cite{AM}. 
Here we present those ideas of the proof that are relevant to understanding the rest of this paper. The key 
observation is that any two concrete diagrams in $S^3_{min}[F,F']$ with $|F|+|F'|$ components are 
necessarily $\strhet$-related (in fact, $\striso$-related) and have equal connectivities. (Both statements follow 
from the fact that all components in any such concrete diagram are disks, each disk bounded by exactly one
loop in either $F$ or $F'$.) The proof then proceeds by induction in $k=|F|+|F'|-n$ where $n$ is the number of 
components of $A$ and $B$. Given $A,B\in S^3_{min}[F,F']$ with $p(A)=p(B)$ and $n<|F|+|F'|$ 
components (\ie, $k>0$), concrete diagrams $A',B'\in S^3_{min}[F,F']$ are constructed such that 
$p(A')=p(B')$, and both have $n+1$ components (\ie, $k$ reduced by one). The induction hypothesis is
that the Theorem is true for $k-1$, so that $A'\strhet B'$. It is then shown that $A$ and $B$ can be 
reconstructed from $A'$ and $B'$ by bridging, in a way that shows that $A$ and $B$ are also $\strhet$-related. 
This shows that the Theorem is true for $k$ if it is true for $k-1$. Being true for $k=0$ (concrete diagrams
in $S^3_{min}[F,F']$ with $|F|+|F'|$ components), it follows that it is always true.

\proposition{finpif}{
For any $F$, $\Pi_F$ is a finite group.}

{\sl Proof: } It was established in Proposition \idclasses\ that $\Pi_F$ is a group. Because $F$
is always a finite set, the partition set $\wp(F\dot\cup F)$ is also finite. Hence, by Theorem
\bigtheorem, there is a finite number of $\strhet$-classes in $[{\mathfrak I}_F]_\het$. \block
\nlabel\finpif{\anylabel} 

\section{Composition of $\het$-classes}
\setcounter{equation}{0}
{\global\propnumber=0}
{\global\defnumber=0}
{\global\lemmanumber=0}
{\global\theonumber=0}
{\global\notnumber=0}
\indent

Consequent to Proposition \hettostrhet, there are fewer $\het$-classes than
$\strhet$-classes in $S^3_{min}[F,F']$. We regard the extra symmetry $\het/\strhet$ as the 
higher-dimensional counterpart of order-preserving displacements of the endpoints of Temperley-Lieb
diagrams along the upper or lower edges. For that reason we regard $\het$-classes as the
natural generalisation of the notion of Temperley-Lieb diagram to higher dimensions, and they will be
the main object of study in the rest of this work.

Our goal, then, is to define a composition of $\het$-classes by which each $\S_\het^3[F,F]$ becomes a 
finite-dimensional, associative algebra with unit. 

\proposition{shmu}{
Let $A,A',B,B'\in S^3_{min}$ and $A\strhet A'$ and $B\strhet B'$. If $A$ and $B$ are 
$\circ$-composable, then $\mu(A\circ B)=\mu(A'\circ B')$.}

{\sl Proof:} From Lemma \shsh, $\mapd(A\circ B)=\mapd(A'\circ B')$. Therefore we only need to
prove that $b(A\circ B)=b(A'\circ B')$ and $g(A\circ B)=g(A'\circ B')$. 
By Lemma \shsh\ we know that $(A\circ B)\strhet(A'\circ B')$, and then by Lemma \bubhet\
$b(A\circ B)=b(A'\circ B')$. By Theorem \bigtheorem\
we know that $p(A)=p(A')$ and $p(B)=p(B')$, hence $p(A\circ B)=p(A'\circ B')$ and,
in particular, $A\circ B$ and $A'\circ B'$ have equal number of non-bubble 
components. Hence $|A\circ B|=|A'\circ B'|$. But it follows from
Eq. \ref{eq:genusformula} that $g(A\circ B)$ depends only on the 
number of components of $A\circ B$ if $A$ and $B$ are minimal. Therefore  
$g(A\circ B)=g(A'\circ B')$. \block
\nlabel\shmu{\anylabel} 

This Proposition can be extended to show that, if $A_i,A_i'\in S^3_{min}[F_i,F_{i+1}]$ with
$A_i\strhet A_i'$ for $i=1,\ldots,n$, then
\beq
\mu(A_1\circ\cdots\circ A_n)=\mu(A'_1\circ\cdots\circ A'_n).
\label{eq:muconc}
\eeq

\definition{complete}{
A subset of $[{ \mathfrak I}_F]_\het$ is {\rm complete} if it consists of exactly one element from 
each $\strhet$-class. }
\nlabel\complete{\anylabel} 

\lemma{redcomp}{
Let $A,A'\in S^d_{min}[F',F]$ and $B,B'\in S_{min}^d[F,F'']$, with $A\het A'$ and 
$B\het B'$. Let $X,X'$ be complete subsets of $[{ \mathfrak I}_F]_\het$. Then 
\bearn
\mu(A\circ \sigma_X\circ B)=\mu(A'\circ \sigma_{X'}\circ B').
\eearn}
\nlabel\redcomp{\anylabel} 
{\sl Proof:} By Proposition \finpif, $X$ and $X'$ are finite sets and $\sigma_X$ and $\sigma_{X'}$ are defined.
By Proposition \hettostrhet\ there are $L_A\het {\mathfrak I}_{F'}$, 
$R_A\het {\mathfrak I}_F$ such that $A\strhet(L_A\circ A'\circ R_A)$, and similarly
$L_B\het {\mathfrak I}_F$, $R_B\het {\mathfrak I}_{F''}$ such that
$B\strhet(L_B\circ B' \circ R_B)$. Then, for each $Y\in\sigma_X$, Eq. (\ref{eq:muconc}) gives
$\mu(A\circ Y\circ B)=\mu(L_A\circ A'\circ R_A\circ Y \circ L_B\circ B' \circ R_B)$.
But, by completeness of $X$ and Proposition \idclasses\ there exists exactly one $Y'\in X'$
such that $(R_A\circ Y\circ L_B)\strhet Y'$, so $\mu(A\circ Y\circ B)=\mu(L_A\circ A'\circ Y'\circ B' \circ R_B)$.
Finally, $(L_A\circ A'\circ Y'\circ B' \circ R_B)\het (A'\circ Y'\circ B')$ and both have equal
number of bubbles and handles, so $\mu(A\circ Y\circ B)=\mu(A'\circ Y'\circ B')$.
The Lemma follows by summing over $Y\in X$. \block

That is, $\mu(A\circ \sigma_X\circ B)$ depends only on the $\het$-classes of $A$ and $B$ and not
on the choice of $X$. Therefore we have a well-defined composition of $\het$-classes into $\K\S^3_\het$:
\beq
[A]_\het \cdot' [B]_\het=\mu(A\circ \sigma_X\circ B),
\eeq
If $|X|$ has an inverse in $\K$ we define a new composition rule that has $[{\mathfrak I}_F]_\het$ 
as unit:
\beq
[A]_\het \cdot [B]_\het={1\over |X|}\mu(A\circ \sigma_X\circ B),
\label{eq:therule}
\eeq
As was done in Section 3 with $\strhet$-classes, the set $\S^3_\het[F,F']$ of $\het$-classes in $S^3[F,F']$
can be interpreted as the set of morphisms with object set $S^2_{[]}$ and composition rule 
given by \ref{eq:therule}. In the next subsection we show that this composition rule is associative. Then,

\theorem{cath}{
The triple $\C_\het=( S^2_{[]}\>,\> \K\S^3_\het[-,-], \>\cdot\>)$ 
defines a category which we will be called the {\it heterotopy category}. }\block
\nlabel\cath{\anylabel} 

The only non-trivial step in the proof of this Theorem is associativity:

\proposition{assoc}{
Let $[A]_\het, [B]_\het, [C]_\het\in\S^3$. Then 
\bearn
([A]_\het \cdot [B]_\het)\cdot [C]_\het= [A]_\het \cdot ([B]_\het\cdot [C]_\het).
\eearn}
{\sl Proof:} Let $X$ be a complete subset of $[{\mathfrak I}_F]_\het$. For each $Y\in X$ define 
$[E_Y]=\mapd(A\circ Y\circ B)$ and $[H_Y]=\mapd(B\circ Y'\circ C)$. Then
\bearn
& &([A]_\het \cdot [B]_\het)\cdot [C]_\het =\\ 
&&\> ={1\over |X|^2}\sum_{Y,Y'\in X}p^{g(A\circ Y\circ B)}q^{b(A\circ Y\circ B)}
p^{g(E_Y\circ Y'\circ C)}q^{b(E_Y\circ Y'\circ C)}\mapd(E_Y\circ Y'\circ C),\\
&&{[A]_\het} \cdot ([B]_\het \cdot [C]_\het) =\\
&&\> ={1\over |X|^2}\sum_{Y,Y'\in X}p^{g(A\circ Y\circ H_{Y'})}q^{b(A\circ Y\circ H_{Y'})}
p^{g(B\circ Y'\circ C)}q^{b(B\circ Y'\circ C)}\mapd(A\circ Y\circ H_{Y'}).
\label{eq:assone}
\eearn
Because both $E_Y\circ Y'\circ C$ and $A\circ Y\circ H_{Y'}$ are in 
$\mapd(A\circ Y\circ B\circ Y'\circ C)$ we have 
$\mapd(E_Y\circ Y'\circ C)=\mapd(A\circ Y\circ H_{Y'})$. Therefore we only need to prove that
\begin{displaymath}
\begin{array}{c}
g(A\circ Y\circ B)+g(E_Y\circ Y'\circ C) =  g(A\circ Y\circ H_{Y'})+g(B\circ Y'\circ C),\\
b(A\circ Y\circ B)+b(E_Y\circ Y'\circ C) =  b(A\circ Y\circ H_{Y'})+b(B\circ Y'\circ C).
\end{array}
\end{displaymath}
Those equations follow from Eqs.(\ref{eq:bandg}) in page \pageref{eq:bandg}. \block
\nlabel\assoc{\anylabel} 

The rule (\ref{eq:therule}) does not provide a practical means of computing the 
composition of two given $\het$-classes, except in the simplest cases where representative concrete 
diagrams can be concatenated by hand. We address this question next.

\section{Computing the multiplication table}
\setcounter{equation}{0}
{\global\propnumber=0}
{\global\defnumber=0}
{\global\lemmanumber=0}
{\global\theonumber=0}
{\global\notnumber=0}

In order to generate examples, and to begin analysing the structure of the algebra, we need to be able to
compute compositions efficiently. If we wished to compute the multiplication table for the algebra
$S^3_\het[F,F]$ with $F$ two concentric circles, we could use as a concrete basis the concrete diagrams in 
Figure 2, concatenate them by hand and finally apply the reduction map $\mu$. This would result in the 
following multiplication table, in which $D_1,\cdots,D_9$ refers to the concrete diagrams shown in Figure 2, 
numbered from left to right.

{\scriptsize
\begin{displaymath}
\vbox{
{\offinterlineskip
\tabskip=0pt
\halign{
\hfil # \hfil & \vrule height2.75ex depth1.25ex width 0.6pt #
& \hskip 5pt\hfil # \hfil & \hfil # \hfil & \hfil # \hfil & \hfil # \hfil
& \hfil # \hfil & \hfil # \hfil & \hfil # \hfil & \hfil # \hfil
& \hfil # \hfil \cr
row $\times$ col
& & $D_1$ & $D_2$ & $D_3$ & $D_4$ & $D_5$ & $D_6$ & $D_7$ & $D_8$ & $D_9$ \cr
\multispan{11}\hrulefill\cr
$D_1$ & & $pD_1$ & $pD_2$ & $D_1$ & $pD_4$ & $D_1$ & $pD_6$ & $D_2$ & $D_4$
& $D_6$\cr
$D_2$ & & $D_1$ & $D_2$ & $qD_1$ & $D_4$ & $D_2$ & $D_6$ & $qD_2$ & $qD_4$
& $qD_6$\cr
$D_3$ & & $pD_3$ & $pD_7$ & $D_3$ & $pD_8$ & $D_3$ & $pD_9$ & $D_7$ & $D_8$
& $D_9$\cr
$D_4$ & & $pD_4$ & $pD_6$ & $D_4$ & $pqD_4$ & $D_4$ & $pqD_6$ & $D_6$ & $qD_4$
& $qD_6$\cr
$D_5$ & & $D_1$ & $D_2$ & $D_3$ & $D_4$ & $D_5$ & $D_6$ & $D_7$ & $D_8$
& $D_9$\cr
$D_6$ & & $D_4$ & $D_6$ & $qD_4$ & $qD_4$ & $D_6$ & $qD_6$ & $qD_6$ & $q^2D_4$
& $q^2D_6$\cr
$D_7$ & & $D_3$ & $D_7$ & $qD_3$ & $D_8$ & $D_7$ & $D_9$ & $qD_7$ & $qD_8$
& $qD_9$\cr
$D_8$ & & $pD_8$ & $pD_9$ & $D_8$ & $pqD_8$ & $D_8$ & $pqD_9$ & $D_9$ & $qD_8$
& $qD_9$\cr
$D_9$ & & $D_8$ & $D_9$ & $qD_8$ & $qD_8$ & $D_9$ & $qD_9$ & $qD_9$ & $q^2D_8$
& $q^2D_9$\cr}}
}
\end{displaymath}
}%

We now present a more efficient method of computing multiplication tables which makes use of the bijection 
established in theorem \bigtheorem\
between $\strhet$-classes in $S^3_{min}[F,F']$ and those partitions of $F\dot\cup F'$ which are connectivities 
of concrete diagrams in $S^3_{min}[F,F']$. A convenient way to represent a partition of $F\dot\cup F'$ 
for this purpose is by means of coloured graphs:

\definition{coldef}{
Let $G$ be a graph and $C$ a set. A {\rm colouring} of $G$ by $C$ is a map from the edge set of $G$
to $C$. Write $G^C$ for the set of all colourings of $G$ by $C$.}
\nlabel\coldef{\anylabel} 

Given $F$, let $\G(F)$ be the rooted undirected graph constructed as follows. The vertex set is the set of
connected components of $\R^2\setminus F$ (``regions''), the root being the vertex associated to the
unbounded region; there is an edge between two vertices if there is a component in $F$ which is a boundary
between the corresponding regions. 

We may associate a rooted tree to the graph $\G(F)$ by forgetting the labels on all the vertices
except the root. Now consider a concrete diagram $D\in S_{min}^3[F,F']$ and an injective map $f$ 
with domain the set of component of $D$. We will say that component $d$ has ``colour'' 
$f(d)$. Given the pair $(D,f)$ we define a colouring $\phi(D,f)$ of the edges of the ordered pair of graphs 
$(\G(F),\G(F'))$ as follows: if $l\in F\dot\cup F'$ is in the boundary of component $d$ in $D$, then the colour
of the edge associated to $l$ is $f(d)$.

We will regard $(\G(F),\G(F'))$ as the single tree $\G(F\dot\cup F')$ by identifying the roots.
Then $\phi(D,f)$ defines a partition of the set of edges of $\G(F\dot\cup F')$ which corresponds
to the partition $p(D)$ of the set of loops of $F\dot\cup F'$. By Theorem \bigtheorem, each such colouring
of $\G(F\dot\cup F')$ corresponds to a $\strhet$-class in $\S^3_\strhet[F,F']$.

\begin{figure}
\centering
\leavevmode
\epsfbox{ 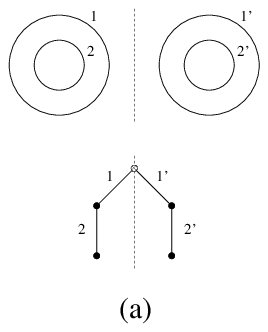   }\hskip 1cm
\epsfbox{ 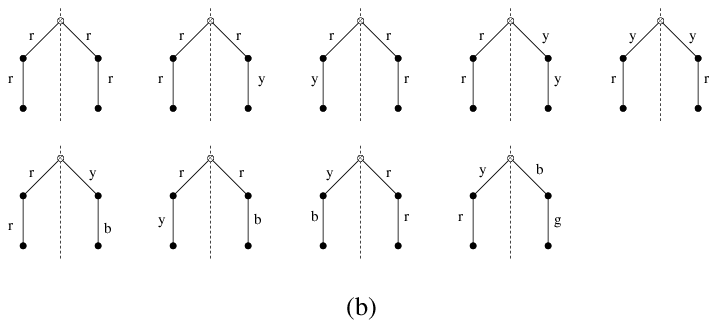  }
\caption{(a) Construction of $\G(F\dot\cup F)$ from Figure 2. (b) Admissible colourings}
\label{fig:tb}
\end{figure}

\definition{admisscol}{
For any two edges $e$ and $e'$ in a tree let $ch(e,e')$ be the chain of edges connecting $e$ to $e'$ in the 
tree (excluding $e$ and $e'$). An element of $\G(F\dot\cup F')^C$ is {\rm admissible} iff for every pair of 
same-coloured edges $e$ and $e'$, either there is another edge in $ch(e,e')$ of the same colour, or else 
every colour in $ch(e,e')$ appears an even number of times.}
\nlabel\admisscol{\anylabel} 

\proposition{realis}{
The image under $\phi$ of the set of $n$-component elements of $S_{min}^3[F,F']$ is
the set of admissible colourings in $\G(F\dot\cup F')^{\{1,2,\cdots,n\}}$. }\block
\nlabel\colrule{\anylabel} 

We refer to \cite{AM} for a proof. This Proposition establishes a correspondence between 
admissibly-coloured graphs and $\strhet$-classes. Then a $\het$-class can be represented
by the coloured graph of any of the $\strhet$-classes of which it consists. 

The correspondence between $\het$-classes and coloured graphs gives us a practical way
of computing compositions of $\het$-classes. Let $A\in S^3_{min}[F',F]$ and $B\in S^3_{min}[F,F'']$ and
$f$ a map colouring the components of $A$ and $B$ so that no colour appears in both $A$ and $B$. To
compute $[A]_\het \cdot [B]_\het$ we first draw the coloured trees $\phi(A,f)$ and 
$\phi(B,f)$. If the group $\Pi_F$ has only one element (the identity) there is a well-defined 
correspondence between the $F$-edges of $\phi(A,f)$ and the $F$-edges of $\phi(B,f)$. 
We then identify the colours of every pair of edges that are in correspondence, and propagate this
identification to the $F'$-edges of $\phi(A,f)$ and the $F''$-edges of $\phi(B,f)$. The element of 
$\G(F'\dot\cup F'')^C$ obtained by joining $\phi(A,f)$ and $\phi(B,f)$ at the root after the identification 
and propagation of colours corresponds to $[A]_\het \cdot [B]_\het$. Therefore composition of
$\het$-classes coincides with the partition algebra composition of the connectivities if $\Pi_F$ is
trivial. This is not so if the group $\Pi_F$ has $n>1$ elements. Then there are $n$ ways in which 
the $F$-edges of $\phi(A,f)$ and the $F$-edges of $\phi(B,f)$ can be put in correspondence. The insertion 
of a complete set in Eq. (\ref{eq:therule}) corresponds to defining the composition to be the uniformly 
weighted sum of the $n$ possible outcomes. 

The power of $q$ in $[A]_\het\cdot[B]_\het$ is the number of colours 
after identification which do not propagate to the $F'$-edges of $\phi(A,f)$ or the $F'$-edges of $\phi(B,f)$. 
If the number of colours in $\phi(A,f)$ and $\phi(B,f)$ before and after identification are respectively $C_b$ and
$C_a$, then by Eq. (\ref{eq:genusformula}) the power of $p$ in $[A]_\het\cdot[B]_\het$ is $C_a-C_b+|F|$.

For example, consider again the concrete diagrams shown in Figure 2. Note first that $\Pi_F=1$, as the two 
concentric loops in $F$ cannot be interchanged by an isotopy. There are nine admissible colourings. Let 
$D_1,D_2,\cdots,D_9$ be the $\het$-classes defined by those admissible colourings, in the order shown in 
the right-hand side of Figure 3 (colours are indicated by letters r, y, b and g). The composition table for 
these heterotopy classes is then the one shown at the beginning of this section.

\begin{figure}
\centering
\leavevmode
\epsfbox{ 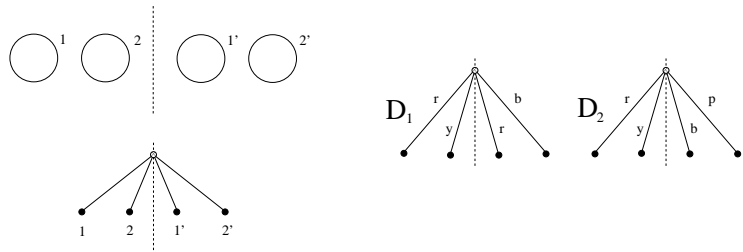  }
\caption{Boundary configuration $(G,G)$ with $\Pi_G=\Z_2$, and two admissible colourings 
$D_1$ and $D_2$.}
\label{fig:tc}
\end{figure}

This example is atypical in that the loop configuration $(F,F)$ has $\Pi_F=1$. Let us now consider the
boundary configuration shown in Figure 4. There are now two different ways in which the coloured edges of 
two concrete diagrams can be put in correspondence. 
Taking the coloured trees shown in Figure 4, we find that the two elements of $\Pi_G$ give rise to
two different contributions to the composition $D_1\cdot D_1$, one of which is $qD_1$ and the other $D_2$.
Therefore we find that $D_1\cdot D_1={1\over 2}(qD_1+D_2)$.

\medskip

A more detailed analysis of the algebras defined here is given in \cite{AM}, together with a preliminary analysis 
of their representation theories, which are intriguingly much richer than the original Temperley-Lieb algebra
itself.

\end{document}